\documentclass[twocolumn,showpacs,preprintnumbers,groupedaddress,english]{revtex4-1}
\usepackage{amsmath}
\usepackage{amssymb}
\usepackage{dcolumn}
\usepackage{graphicx}
\usepackage{longtable}
\usepackage{physics}
\usepackage{bm}
\usepackage{color}
\usepackage{epsfig}
\usepackage{hyperref}
\usepackage[T1]{fontenc}
\usepackage{ulem}
\usepackage[utf8]{inputenc}

\hypersetup{colorlinks=true, linkcolor=blue, citecolor=blue, urlcolor=blue}

\begin{document}

\title{Nonlinear Hall Effects induced by Berry Curvature Dipole in CuPb$_9$(PO$_4$)$_6$O} 
\author{Bishnu Karki}
\affiliation{Department of Physics and Texas Center for Superconductivity, University of Houston, Houston, TX 77204}
\author{Kai Chen }
\affiliation{Department of Physics and Texas Center for Superconductivity, University of Houston, Houston, TX 77204}
\author{Pavan Hosur}
\affiliation{Department of Physics and Texas Center for Superconductivity, University of Houston, Houston, TX 77204}

\date{\today}
\begin{abstract}
The nonlinear Hall effect (NLHE), an emergent response in systems with broken inversion symmetry, provides a powerful tool for probing topological transport properties. In this context, we investigate copper-substituted lead apatite (LK-99), a material that initially garnered attention for its controversial claim of room-temperature superconductivity. Despite the unresolved nature of its superconducting properties, LK-99's unique electronic structure characterized by flat bands near the Fermi level and broken inversion symmetry makes it a promising candidate for exploring Berry curvature-driven phenomena, such as the NLHE. Using first-principles density functional theory and an augmented tight-binding Hamiltonian model, we investigate LK-99's band topology and transport properties. Our calculations indicate that spin-orbit coupling in LK-99 generates multiple Weyl points near the Fermi level, thereby enhancing the Berry curvature distribution by further splitting the bands. Crucially, the absence of inversion symmetry in LK-99 leads to a net Berry curvature dipole, producing a nonlinear Hall current that scales quadratically with the applied electric field. The nonlinear Hall effect is solely due to the BCD, as the contributions from the Drude weight and quantum metric are zero due to time reversal symmetry. Moreover, we demonstrate that the NLHE in LK-99 can be tuned by varying the direction of the applied electric field, underscoring its potential as a versatile platform for exploring topological transport phenomena and designing next-generation nonlinear electronic devices.
\end{abstract}

%
%
%
%
%

\maketitle

\onecolumngrid
\section{Introduction}

The nonlinear Hall effect (NLHE) is a second-order response in time-reversal-invariant systems, where a transverse voltage is generated in response to the square of an applied electric field \cite{du2021nonlinear,ma2019observation}. Unlike the conventional Hall effect, which relies on an external magnetic field, the NLHE arises intrinsically from the Berry curvature dipole (BCD) a dipolar distribution of Berry curvature in momentum space \cite{ortix2021nonlinear}. This phenomenon is of great interest due to its potential applications in next-generation electronic devices, such as rectifiers and energy harvesters, as well as its ability to probe the topological properties of materials. NLHE offers a powerful probe of inversion symmetry breaking and topological transport, linking electronic structure to Berry curvature. Notably, it enables a transverse voltage without the need for an external magnetic field, in both metals and insulators. This intrinsic, symmetry-driven response holds promise for applications in modulators, photodetectors, and light-emitting diodes, as well as high-sensitivity strain sensors and wireless radio frequency rectifiers \cite{kumar2021room,isobe2020high}. Due to its sensitivity to crystal geometry, the NLHE also provides insight into topological phase transitions and quantum geometric properties, with emerging relevance in energy harvesting and current rectification technologies \cite{zhang2021terahertz,Wang2025,bandyopadhyay2024non,ortix2021nonlinear}. Recent studies have identified materials with low crystal symmetry, such as transition metal dichalcogenides \cite{PhysRevApplied.13.024053, Taghizadeh_2020,ye2023control,zhao2023berry}, Weyl semimetals \cite{PhysRevB.102.245116}, and topological insulators \cite{PhysRevB.103.155415}, as promising candidates for exhibiting prominent nonlinear Hall responses. 

Among these, Weyl semimetals have emerged as particularly fascinating systems for investigating the nonlinear Hall effect. As topological materials with broken inversion or time-reversal symmetry, Weyl semimetals feature Berry curvature concentrated around Weyl nodes—band-touching points in momentum space \cite{CRPHYS_2013__14_9-10_857_0}. The unique electronic structure of Weyl semimetals, combined with their low crystal symmetries, makes them ideal platforms for realizing a substantial BCD\cite{Roy_2022,bandyopadhyay2024non, PhysRevLett.125.046402, PhysRevB.98.121109,zhang2018berry}. Furthermore, the tunability of Weyl semimetals via external perturbations, such as strain, electric fields, or chemical doping, provides a versatile route to engineer and optimize their nonlinear Hall response \cite{singh2020engineering,PhysRevB.105.125138}. These properties not only deepen our understanding of topological transport phenomena but also position Weyl semimetals as candidates for next-generation electronic and optoelectronic devices.

In this context, LK-99, a copper-substituted lead apatite (CuPb$_9$(PO$_4$)$_6$O), presents an compelling case. Initially proposed as a room-temperature ambient-pressure superconductor, LK-99 generated significant excitement in the scientific community \cite{Lee1, Lee2}. The claim was supported by experimental evidence, including zero resistivity, the critical temperature, and the Meissner effect. However, repeated attempts to reproduce these results were unsuccessful, raising doubts about the superconductivity claim\cite{kumar2023synthesis,liu2023semiconducting,timokhin2023synthesis,thakur2023synthesis,singh2023experimental,guo2023ferromagnetic,wu2023successful,doi:10.1021/acs.jpcc.3c05684, ZHU20234401, Habamahoro_2024}. Despite this setback, LK-99 remains a material of interest due to its unique electronic structure, particularly the presence of flat bands near the Fermi level with multiple degeneracies as reported by the various first-principle based calculations \cite{bai2023semiconductivity, PhysRevB.109.144515,oh2023s, PhysRevB.109.L100504,
griffin2023origin,kurleto2023pb,mao2023wannier,shen2024phase,paudyal2023implications,liu2024different,chen2024berry}.

These flat bands, combined with inversion symmetry breaking and spin-orbit coupling (SOC) effects due to the heavy Pb atoms, suggest that LK-99 may exhibit topological features akin to those observed in Weyl semimetals. First-principle calculations indicate that SOC can split the flat bands, leading to Weyl-like crossings near the Fermi level. This electronic structure creates conditions for an unequal distribution of Berry curvature, making LK-99 a promising candidate for studying nonlinear Hall effects. The ability to manipulate these features through strain, chemical doping, or electric fields further enhances its potential for exploring topological transport phenomena and designing nonlinear electronic devices. Recent advancements in Weyl physics and correlated metallic states further emphasize the relevance of studying the unique electronic structure of LK-99 within this framework\cite{PhysRevB.109.085103,PhysRevMaterials.8.014201,zhou2023cu,kim2024non}. Additionally, the successful experimental synthesis and confirmation of dynamic stability through phonon band calculations reinforce the importance of investigating these properties in greater detail \cite{kim2024dynamical,paudyal2023implications}.

In this study, we investigate the nonlinear Hall effects in LK-99 using first-principle calculations, focusing on the role of inversion symmetry breaking and Berry curvature distribution. The lack of inversion symmetry leads to an asymmetric distribution of Berry curvature, resulting in a nonzero Berry curvature dipole (BCD). Notably, the BCD exhibits peaks at the locations of Weyl points, which play a crucial role in generating the nonlinear Hall current. For the confirmation of the origin of the non linear Hall current we also calculated the contribution from the quantum mteric (QM) dipole, as the contribution from quantum metric is prominent in the systems that breaks both the time reveral and inversion symmetry like CuMnAs \cite{wang2021intrinsic}, MnBi$_2$Te$_4$\cite{wang2023quantum,gao2023quantum} and some altermagnets \cite{fang2024quantum}. Furthermore, the nonlinear Hall current can be tuned by altering the direction of the applied electric field, offering a means to control topological transport behavior. By drawing parallels with Weyl semimetals, we aim to provide deeper insight into LK-99’s topological and transport properties.

The structure of this paper is organized as follows. Section II provides an overview of the crystal structure and the methodologies utilized in this study. Section III details the results and conclusions, focusing on the electronic and topological properties examined through density functional theory (DFT) and an enhanced tight-binding model, along with an analysis of nonlinear Hall effects.

\section{Crystal structure and Methods}
\begin{figure}[t]
   \centering
   \includegraphics[width=1\columnwidth]{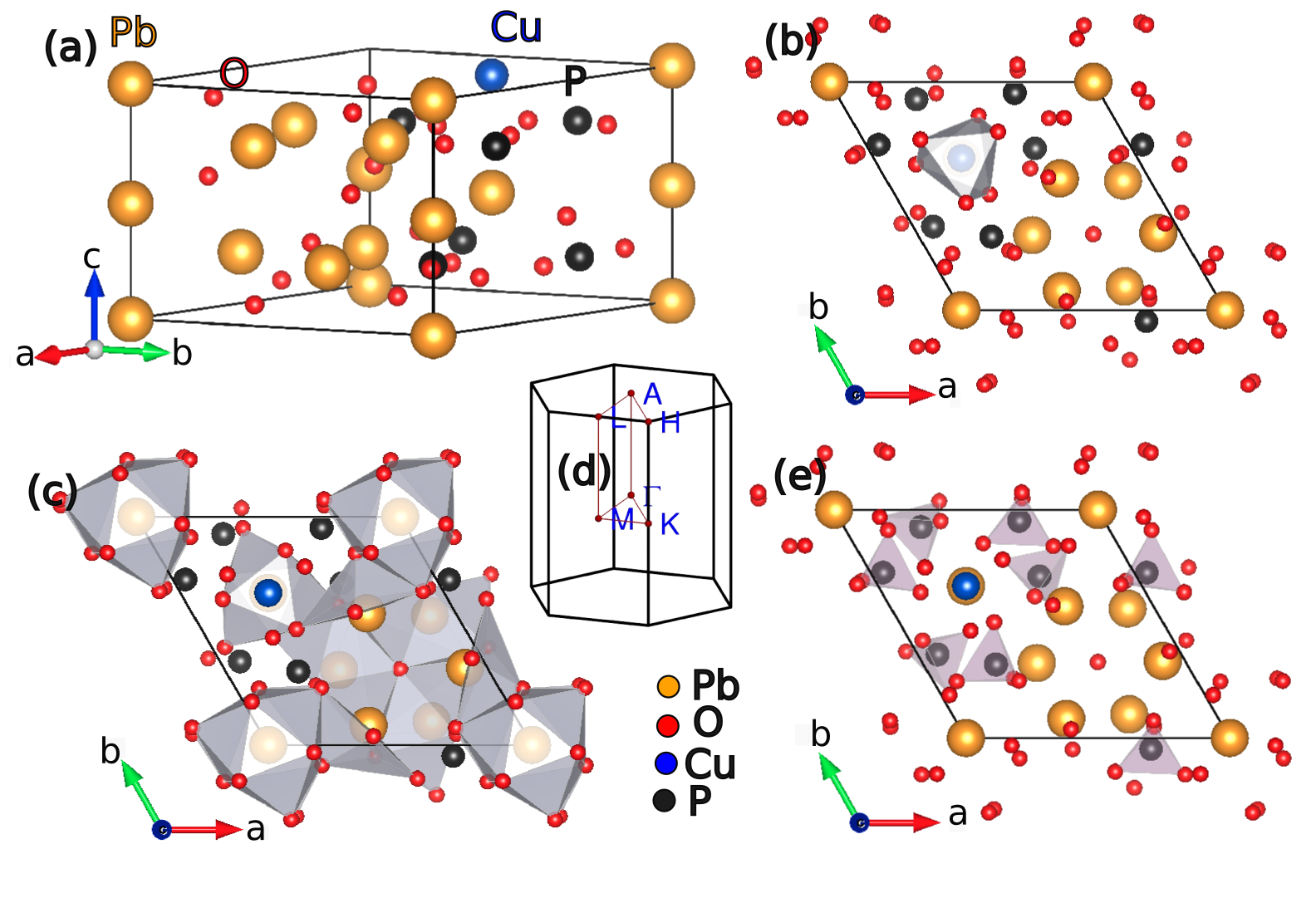}
   \caption{(a) Crystal Structure of LK-99 consisting of oxygen (red), lead (yellow), phosphorus (black) and copper (blue) atoms. (b) Structure seen in a-b plane with Cu forming polyhedra with O. (c) Structure seen in a-b plane with Pb forming polyhedra with O. (d) Brilloun zone (BZ) with high symmmetry points. The chosen points are $\Gamma$(0,0,0), M(0.577,0,0), K(0.577,1/3,0), A(0,0,0.688), L(0.577,0,0.668) and H(0,0,0.688) in the units of ($2\pi/a$, $2\pi/b$, $2\pi/c$). (e) Structure seen in a-b plane with P forming tetrahedra with O.}
    \label{crystal}
  \end{figure}

The original crystal structure of lead hydroxyapatite Pb$_{10}$(PO$_4$)$_6$(OH)$_2$ was initially determined through powder X-ray diffraction by Bruckner et al. in 1995 \cite{bruckner1995crystal}. This structure is categorized under space group 176 (P63/m), which features an inversion center. There are two sites (Pb(I) and Pb(II)) for the Pb atom, which forms a hexagonal crystal sublattice. As one of the Pb(I) sites is replaced by Cu, the substituted Cu is not mapped to itself under inversion. Moreover, in the model we are using, the fractional site occupancy is removed for the Oxygen and hydroxide groups, lowering the symmetry to 143. Thus the structure is non-centrosymmetric and based on a trigonal Bravais lattice. The structure forms a complex network with central atoms Cu, P, and Pb forming polyhedra with O as shown in Fig.\ref{crystal}. The selected lattice parameters are as follows: a = b = 9.80 \AA, c = 7.34 \AA, and $\alpha$ = $\beta$ = 90$^{\circ}$, while $\gamma$ = 120$^{\circ}$.
We employed the full potential local orbital (FPLO) code version 22.00, as outlined in the work by Koepernik and Eschrig in 
1999 \cite{koepernik1999fplo}, to carry out calculations using density functional theory (DFT). The Perdew, Burke, and Ernzerhof (PBE-96) parametrization of the generalized gradient approximation (GGA) was utilized to compute the exchange and correlation 
energy \cite{perdew1996generalized}. Our approach involved self-consistent calculations conducted in the four-component fully relativistic mode of FPLO. The basis sets encompassed valence states for Cu [3s, 3p, 4s, 5s, 3d, 4d, 5d], Pb [5s, 5p, 5d, 6s, 7s, 6d, 6p, 7p],  P [2s, 2p, 3s, 4s, 3p, 4p, 3d]  and  O [1s, 2s, 3s,2p, 3p, 3d]. For Brillouin zone sampling, we employed a $k$-mesh subdivision of $8 \times 8 \times 10$. The energy convergence criteria is set to be 10$^{-5}$ Ha and the charge density convergence critera is 10$^{-4}$ e/(Bohr radii)$^3$. Maximally localized Wannier functions were generated from self-consistent, fully relativistic DFT band structures using the PYFPLO module of the FPLO code. The localized Wannier basis consists of Cu [3d$_{yz}$, 3d$_{xz}$] orbitals. The Wannier model was constructed using the same k-mesh as in the self-consistent calculations. The resulting Wannier Hamiltonians were employed to interpolate the band structure on a 200 × 200 × 200 $k$-point grid and to identify Weyl points, utilizing the PYFPLO module for this purpose. To compute BCD  wannierberri package has been utilized \cite{Tsirkin2021}.

\section{{RESULTS AND DISCUSSION}}
\begin{figure}[h]
   \centering
   \includegraphics[width=1\columnwidth]{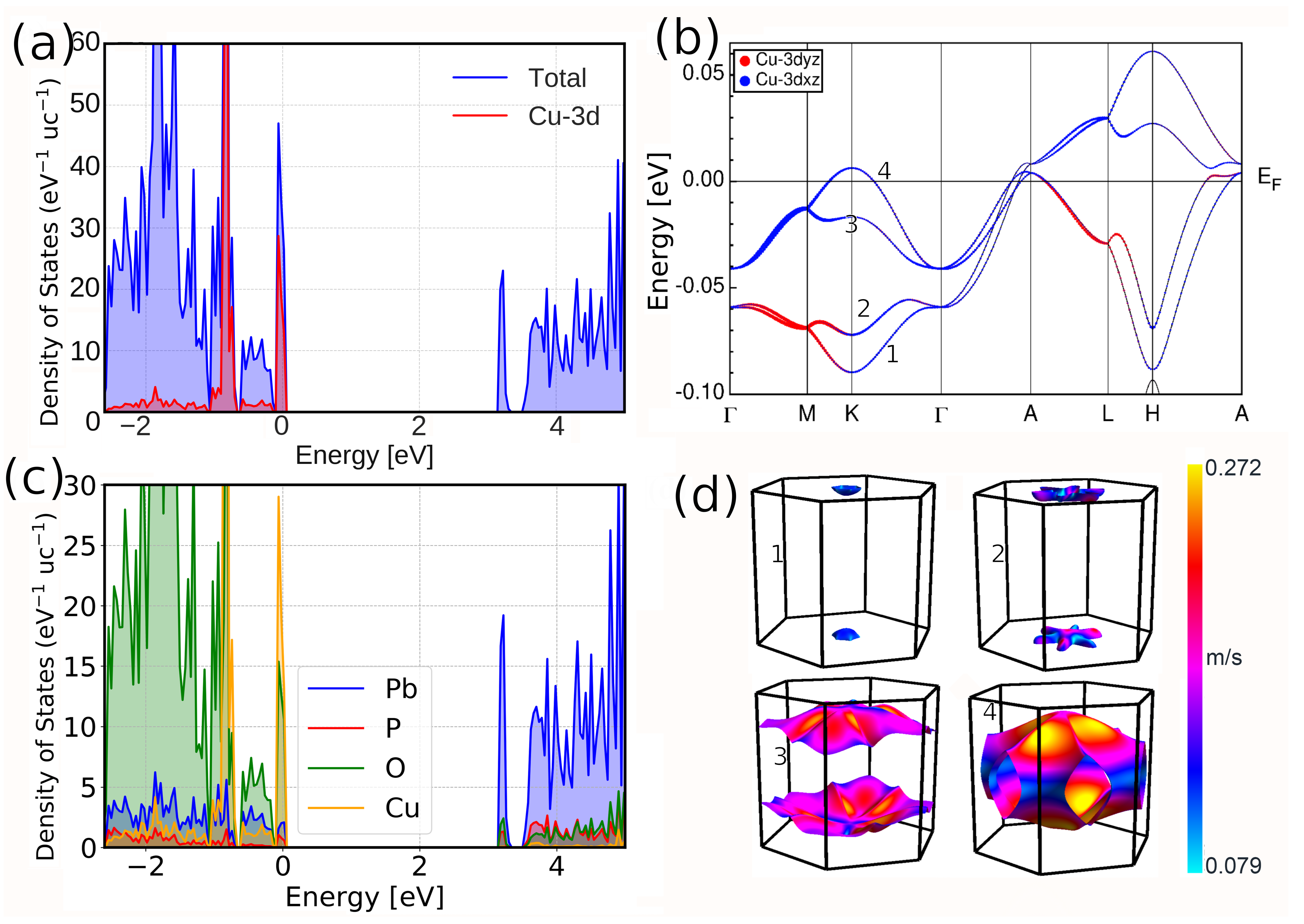}
   \caption{Electronic structure of LK-99. The Fermi level is set to zero energy.(a) Total density of states and partial density of states of Cu- 3d in the units of per eV per unit cell. (b) Fat band analysis that includes the contribution from the Cu-3d$_{yz}$ and Cu-3d$_{xz}$ orbitals. The numbering
    is in the order of the band energy at each k-point.(c) Density of states of the individual atoms of the system. (d) Fermi surface of the four bands that contribute to the Fermi level. The color bar shows the band velocity.}
    \label{dos-band}
  \end{figure}
\subsection{Electronic Features}

 We begin by investigating the electronic structure of LK-99 through an analysis of the density of states and band features. Fig. \ref{dos-band}(a) depicts the total density of states, indicating that the system is metallic due to a prominent peak at the Fermi level. This peak resembles a Van Hove singularity (VHS) and is followed by a relative scarcity of states in the positive energy range up to 3.1 eV. In contrast, the density of states is notably richer below the Fermi level. The presence of the VHS is linked to the existence of flat bands near the Fermi level, primarily contributed by Cu orbitals.
 
 Among the five d orbitals, the most significant contribution to this density is by Cu-3d$_{yz}$ and  Cu-3d$_{xz}$, as indicated by the thickness of the lines in Fig. \ref{dos-band}(b). Here, the thickness of the lines represents the sum of the squares of coefficients associated with the basis states used in the expansion of the Kohn-Sham wavefunction into localized orbitals. We see isolated flat bands near the Fermi level, which are the result of the VHS peak in the density of states in Fig. \ref{dos-band}(b). These flat characteristics of the band have been shown from the minimal model as in Ref \cite{PhysRevB.109.L100504}. This flatness of the bands is more prominent when the Hubbard U potential is taken into account, as done in the Ref \cite{griffin2023origin}. Below these bands, the O-2p orbitals seem to dominate over the Cu-3d orbitals. Most of the transport phenomena of the system are governed by these two bands due to their low energy states; therefore, the rest of our discussion focuses on the features of these bands.

   \begin{figure}[t]
   \includegraphics[width=1\columnwidth]{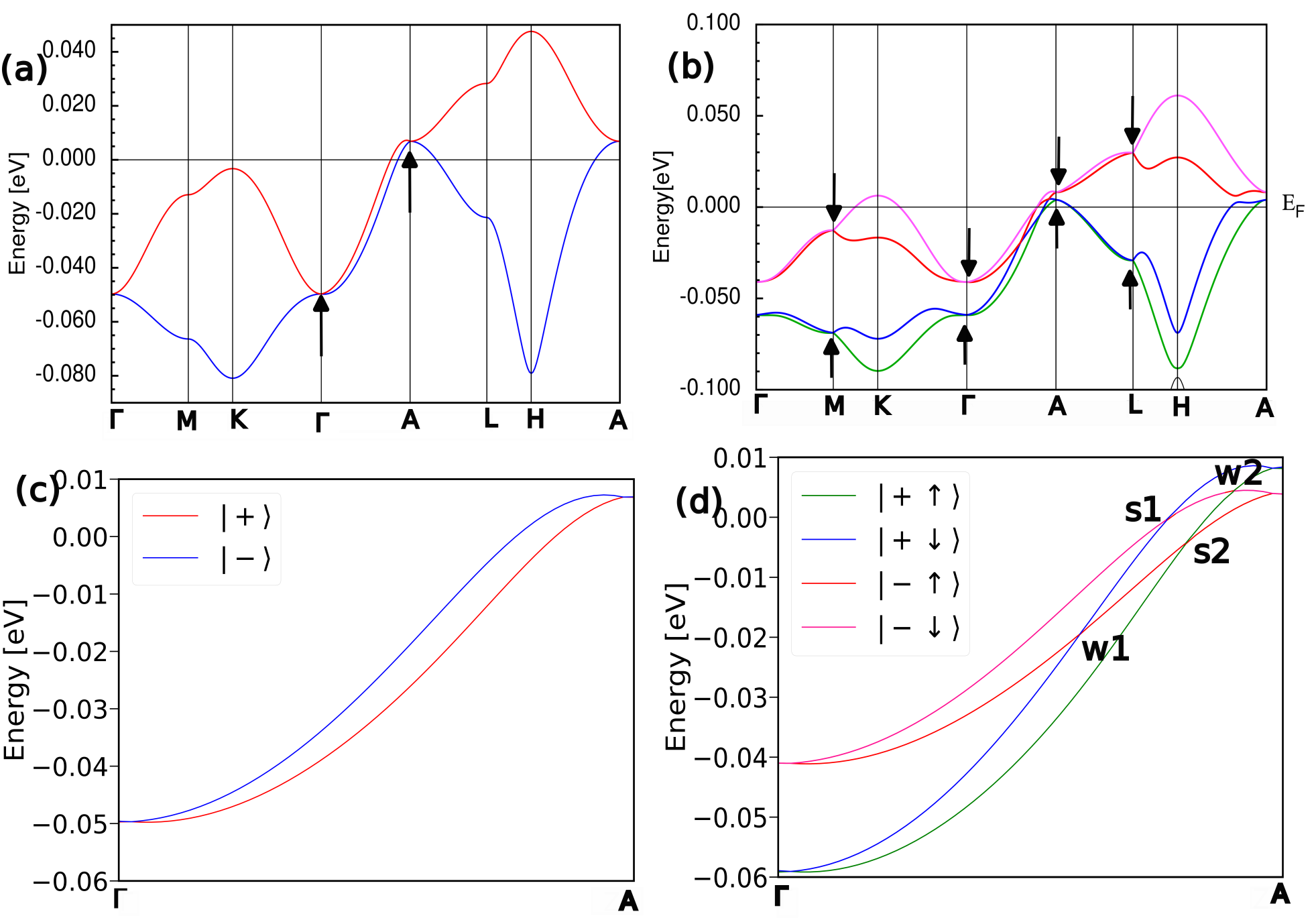}
   \caption{Comparison of electronic band structure with and without SOC. a) Isolated bands near the Fermi level without considering SOC. Arrows highlight double degeneracies at $\Gamma$ and A protected by $C_{3z}$ and spinless time-reversal symmetry. b) Splitting of the bands  after consideration of the case of spin orbit coupling. The arrows in the band indicate the  doubly degenerate points at the high symmetry points. c) Zoomed-in view of the region $\Gamma$-A without SOC d) Zoomed-in view of the region $\Gamma$-A with SOC.}
    \label{fig3}
  \end{figure}

The Fermi surfaces (FSs) plot, which includes the effects of spin-orbit coupling (SOC), is shown in Fig. \ref{dos-band}(d). Four bands contribute to the FS: bands 1, 2, and 3 create flat, electron pocket-like FSs, while the fourth band generates an open FS that spans the entire Brillouin zone (BZ). Incorporating the effects of SOC yields significant changes in the band structure, as demonstrated in Fig. \ref{fig3}(b) and (d). In the absence of SOC, two isolated bands are initially positioned near the Fermi level. There are two degenerate points present at the high symmetry points $\Gamma$ and A, as indicated by the arrows in Fig. \ref{fig3}(a). These points were proposed based on symmetry arguments in tight-binding models \cite{PhysRevMaterials.8.014201, zhou2023cu} and were shown to carry Weyl charge $\pm2$. Upon activation of spin-orbit coupling, these two bands split into four distinct bands. Notably, the introduction of SOC leads to the creation of multiple doubly degenerate points. At the $\Gamma$ point, we see two doubly degenerate points with a small gap in between them. The degeneracies at $\Gamma$ point is between the Kramer's pair \{$\ket{+\uparrow}$ and $\ket{+\downarrow}$ \}, \{$\ket{-\uparrow}$ and $\ket{-\downarrow}$ \}. Other symmetry-protected double degeneracies are also observed at the high symmetry points M, A, and L.

Intriguingly, the high symmetry line $\Gamma$-A experiences distinctive behavior. Four doubly degenerate points (labeled as w1, w2, s1, and s2) emerge when SOC is considered, as depicted in Fig. \ref{fig3}(d). Among these points, w1 is the mixing between the states \{$\ket{+\downarrow}$, $\ket{-\uparrow}$\} and w2 as a crossing point between \{$\ket{+\uparrow}$, $\ket{-\downarrow}$\}. Point s1 signifies the mixing between \{$\ket{+\uparrow}$, $\ket{-\uparrow}$\}, while point s2 represents the crossing between \{$\ket{+\downarrow}$, $\ket{-\downarrow}$\}. Here, we see that w1 and w2 are the result of mixing between different orbitals and different spins, while s1 and s2 are the result of mixing between different orbitals with the same spin states. The nomenclature is based on an analysis of the tight-binding model,
which shows that w1 and w2 are Weyl points while s1 and s2 are points on nodal surfaces. To verify them and study their topological invariants, we modified the Hamiltonian by Zhou and Franz \cite{zhou2023cu}.

\subsection{Comparison with other first-principles computations}

Within the framework of GGA+SOC in a non-magnetic state, our investigation examines the material system's band structure, revealing distinct topological Weyl features. Simultaneously, an alternative study utilizing DFT and a minimal model Hamiltonian without SOC suggests the presence of double Weyl nodes at high symmetry points $\Gamma$ and Z \cite{PhysRevMaterials.8.014201}.
It is noteworthy that our study specifically excludes the incorporation of the Hubbard-U potential. Despite this omission, the semiconductor characteristics of the system are highlighted \cite{bai2023semiconductivity}. Additionally, first-principle calculations still indicate a ferromagnetic behavior within the system, with a magnetic moment of 1$\mu_B$ per unit cell \cite{bai2023ferromagnetic}. Also, there is a claim that the system shows Mott-insulating-like features from dynamic mean field calculations \cite{si2023pb10}.

\subsection{Augmented tight-binding model}

Initially, without considering SOC, our DFT band matches the band from theoretical calculations. But, after consideration of SOC, we see the differences in the band structure. Our results show four doubly degenerate points between $\Gamma$-Z, while their results show only two. The model Hamiltonian with spin-orbit coupling, which they have adapted for band structure:
\begin{equation}
	H = H_0 + H_\text{SOC}		
\end{equation}
Here, $H_0$ is the symmetry based realistic two-band tight binding model and $H_{SOC}$ is the spin orbit coupling term, which has the form,
\begin{equation}
	H_{SOC} = \dfrac{\lambda}{2}\tau_z \otimes \sigma_z
\end{equation}
$\lambda$ being the strength of spin orbit coupling term and $\tau_z$ as orbital degrees of freedom and $\sigma_z$ as spin degrees of freedom. The chosen basis sets are: \\
\big\{$\ket{+,\uparrow}$, $\ket{+,\downarrow}$,$\ket{-,\uparrow}$,$\ket{-1,\downarrow}$\big\}. \\
\begin{figure}[t]
   \centering
   \includegraphics[width=1\columnwidth]{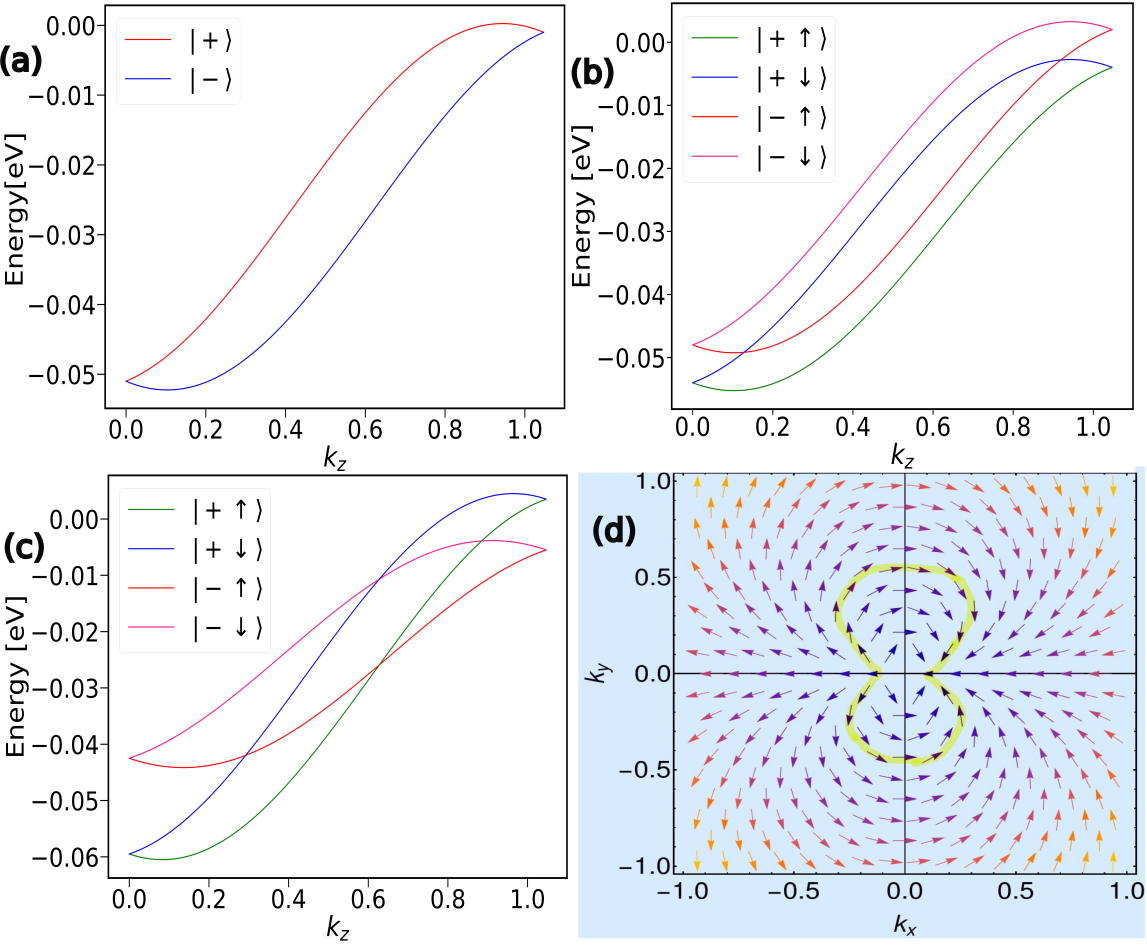}
   \caption{a)  Band structure from the modified model Hamiltonian without inclusion of SOC ($\lambda$=$\lambda'$=0) along $\Gamma$-Z. b) Band structure from the modified model Hamiltonian with inclusion of SOC ($\lambda$ $\neq$0, $\lambda'$=0) along $\Gamma$-Z. c) Band structure from the modified model Hamiltonian with inclusion of SOC ($\lambda$ $\neq$ 0, $\lambda'$ $\neq$0) along $\Gamma$-Z. d)Vector diagram for V=$(f_x,f_y)$ around k$_x$ and k$_y$.}
    \label{fig4}
  \end{figure}

Using this Hamiltonian, we observe only two points of intersection when moving along the $\Gamma$-Z direction, which disagrees with our DFT results, where we see four crossings. To address this discrepancy, we have introduced an additional term in $H_{SOC}$, $\lambda' \cos(k_zc)$, which preserves all the symmetries of the original model, $c$ being the lattice constant in the $z$ direction. Physically, it accounts for the nearest neighbor hopping in the $z$-direction between the Cu-$d$ orbitals. Thus, $H_{SOC}$ now takes the form,
\begin{equation} 
	H_{SOC} = \left(\dfrac{\lambda}{2}+\lambda' \cos(k_zc)\right)\tau_z \otimes \sigma_z
\end{equation}
The additional term in the Hamiltonian switches the eigenstates \{$\ket{+\uparrow}$,$\ket{+\downarrow}$ \} and \{$\ket{-\downarrow}$,$\ket{-\uparrow}$\} while evolving along k$_z$. This generates four doubly degenerated crossings in the band structure. This is presented in Fig. \ref{fig4}(c).

The low energy Hamiltonian around the points w1 and w2 are described by the following Hamiltonian:
\begin{equation}
	H_{w1,w2}=a (k_x^2-k_y^2) \tau_x+2ak_xk_y\tau_y+bk_z\tau_z+ck_z^2\tau_z
\end{equation}
The Hamiltonian takes the form,
\begin{equation}
	H_{w1,w2}=f_x \tau_x+f_y\tau_y+f_z\tau_z
\end{equation}
We define the winding vector V=$(f_x,f_y)$ which is linear in third direction so choosing the dispersion in $k_x-k_y$ plane is sufficient to define the Chern number for the above Hamiltonian (See Appendix D). The vector V=$(f_x,f_y)$ simply makes $\pm$4$\pi$ rotation while coming back to original position in $k_x-k_y$ plane along a loop shown in Fig. \ref{fig4}(d), thus the topological invariant Chern number of the node is given by, 
\begin{equation}
	C_{w1,w2} = 2 \text{sgn(a) sgn(b)}
\end{equation}
Hence, C$_{w1}$=2 and C$_{w2}$=-2 verifies the points w1 and w2 to be double Weyl points. The details of the numerical parameters of a, b, c are presented in supplementary Table \ref{tab1}. 

Similarly, the low energy Hamiltonian at the points s2 and s3 are given by,
\begin{equation}
	H_{s1,s2}=bk_z\tau_z+ck_z^2\tau_z
\end{equation}

The Chern number is not defined for this Hamiltonian, and the degeneracies are just the accidental nodes. These nodes are only protected by $\tau_z$ conservation, and will generically split into  Weyl nodes once terms proportional to $\tau_x$ and $\tau_y$ are added. Thus, from this model, we can say that only w1 and w2 are the real Weyl nodes, while s1 and s2 are accidental crossings. 

\subsection{Non Linear Hall Effects}
\begin{figure}[!h]
   \centering
   \includegraphics[width=1\columnwidth]{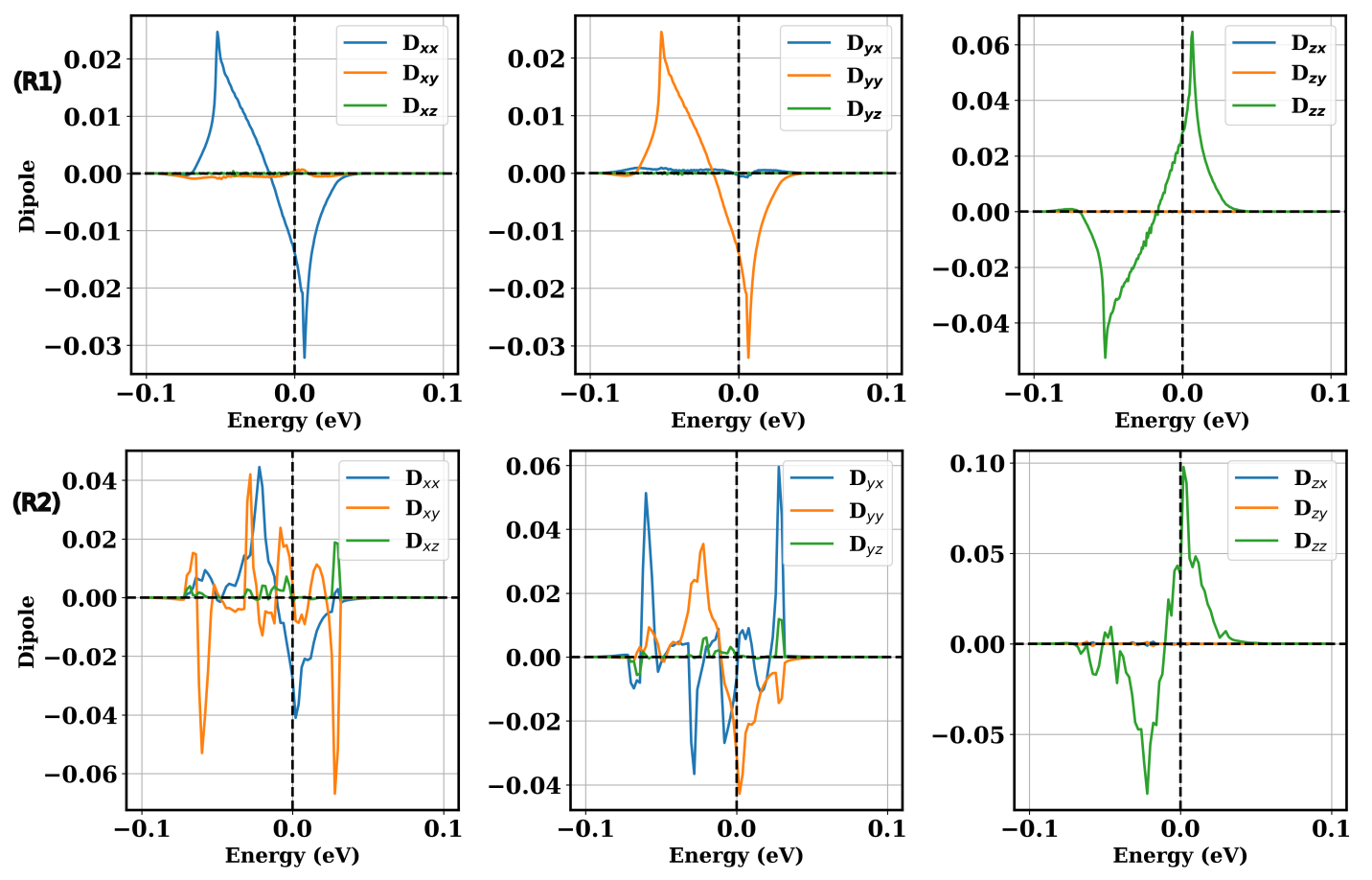}
   \caption{Various components of the Berry curvature dipole plotted against chemical potential. The individual row consists of 3 figures where row1 (R1) case of without SOC  row2 (R2) case of SOC. }
    \label{BCD}
  \end{figure}

To further explore the topological characteristics and transport behavior, we constructed a Wannier Hamiltonian based on the Cu-\(3d_{yz}\) and Cu-\(3d_{zx}\) orbitals. The resulting Wannier functions are well localized, and the computed band structure aligns closely with the DFT band structure (see Appendix, Fig. \ref{fig6}). Since the augmented model Hamiltonian is confined to the \(\Gamma\)-Z region, it effectively captures the features of the entire Brillouin zone (BZ). Without including SOC, the model reveals the presence of two Weyl points at the high-symmetry points \(\Gamma\) and Z. However, incorporating SOC induces band splitting, significantly increasing the total number of Weyl points across the BZ to 60. Detailed information on the Weyl points and their locations can be found in Appendix F. The Weyl points in this system emerge due to the breaking of inversion symmetry. This symmetry breaking leads to an asymmetric distribution of Berry curvature around the Weyl points, resulting in a non-zero Berry BCD, which is the first moment of the Berry curvature. The BCD plays a central role in driving the nonlinear Hall effect under an applied external electric field. The nonlinear Hall current is expressed as:

\begin{equation}
    J_{a}^{2\omega} = \chi_{abc}E_bE_c
\end{equation}
Here, $J$ is the transverse current that is related to the applied electric field by the tensor $\chi$. 
This tensor quantity can be related to the BCD by the following relations\cite{sodemann2015quantum},
 \begin{figure}[!h]
   \centering
   \includegraphics[width=\columnwidth]{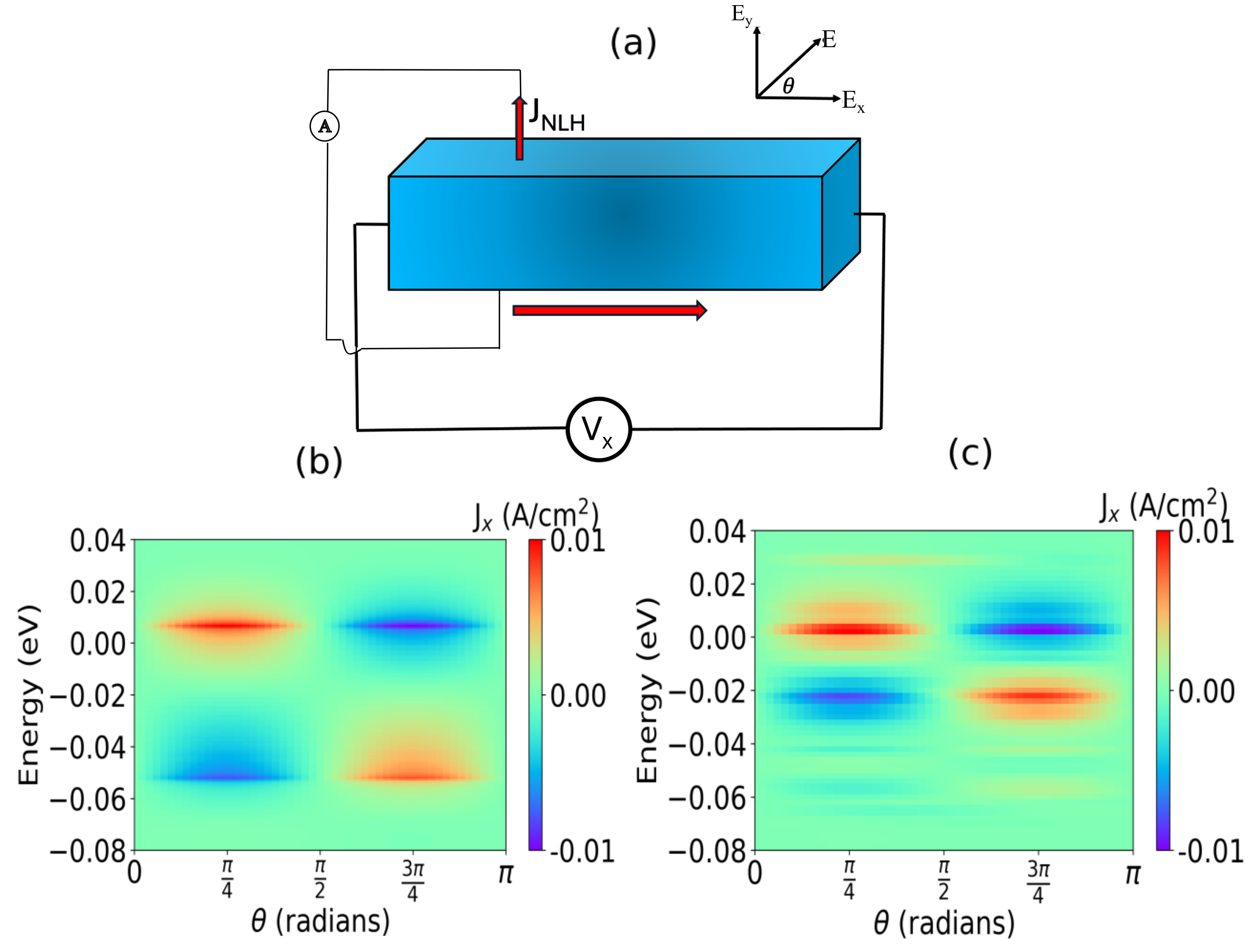}
   \caption{(a) Schematic diagram of non linear Hall effect. Horizontal red arrow is the direction of applied electric field. The vertical red arrow indicates the direction of  non linear Hall current(J$_{NLH}$). (b),(c) Non linear Hall current with respect to the chemical potential and the direction of applied electric field without and with spin orbit coupling respectively. Axis shows the plane of applied electric field and $\theta$ is the direction of applied electric field. Here, considering the static limit $\omega$ $\rightarrow$ 0,  $\tau$ is 1 ps choosen from the expermental Refs. \cite{shekhar2015extremely,arnold2016negative}  and the physically applicable laboratory electric field strength 1 V/cm is considered for the numerical estimation of current.}
    \label{fig5}
  \end{figure}
\begin{equation}
    \chi_{abc} = -\epsilon_{adc}\dfrac{e^3\tau}{2\hbar^2(1+i\omega \tau)}D_{bd}
\end{equation}
where,
\begin{equation}
 D_{bd}(\mu,T)=\int [d\bold{k}]\sum_n \dfrac{\partial E_n}{\partial k_b}\Omega_{n,d}\left(-\dfrac{\partial f(T,\mu,E)}{\partial E}\right)_{E=E_n}   
\end{equation}
 is the BCD , $\Omega$ is the Berry curvature, $\mu$ is the chemical potential, $\tau$ is the relaxation time, $\epsilon$ is the Levi-Civita symbol, $f$ is the Fermi Dirac distribution given by, $f(T,\mu,E)=\dfrac{1}{exp((E-\mu)/kT)+1}$, Integration $[dk]=\dfrac{d^3\bold{k}}{2\pi^3}$, The indices a,b,c are the crystal direction and n is the band index.
We computed the BCD for both the case of SOC and without SOC which are presented in Fig.\ref{BCD}. The various components of the BCD exhibit sharp peaks at the locations of the Weyl points. The first row in Fig. \ref{BCD} highlights peaks at -0.052 eV and 0.0054 eV. In the absence of spin-orbit coupling (SOC), only the diagonal components are non-zero. However, when SOC is activated, we observe the emergence of several sharp peaks, primarily due to the appearance of new Weyl points influenced by SOC effects. Additionally, the band structure exhibits tilts near the nodal points, resulting in asymmetric electronic features. This band tilting is a significant factor contributing to the non-zero BCD \cite{yar2022nonlinear, li2021nonlinear, gao2020second}. The lack of mirror symmetry leads to most of the components being non-zero, adhering to the relation \( D_{ij} - D_{ji} \). Using this calculated BCD, we estimated the nonlinear Hall current of the system. When an electric field of unit magnitude is applied in the \(xy\), \(yz\), and \(zx\) planes, the resulting transverse currents take the following forms: 
\begin{equation}
    J_z = D_{xy}+\dfrac{1}{2}(D_{yy}-D_{xx})\text{sin}2\theta
\end{equation}
\begin{equation}
    J_x = D_{yz}+\dfrac{1}{2}(D_{zz}-D_{yy})\text{sin}2\theta
\end{equation}
\begin{equation}
    J_y = D_{xz}+\dfrac{1}{2}(D_{xx}-D_{zz})\text{sin}2\theta
\end{equation}
 From the BCD values of $D_{yy} ~ D_{xx}$, first term in the eqn (11) dominates making the current unaffaected by the angle whereas the second term domintaes for $J_x$ and $J_y$ making the current to be sinusoidal as illustrated in Fig.\ref{fig5}. The highly intense region of the current corresponds to the location of Weyl points. As from the results we can say that the current can be controlled through chemical doping and changing the direction of applied electric field. Along with BCD contribution to $J$ come from the QM and Drude weight also . The contribution due to the QM and Drude weight is given by \cite{wang2021intrinsic,gao2023quantum},
 \begin{equation}
    \chi^{\alpha\beta\gamma}_{QM}=2e^3\sum_{n,m}^{\epsilon_n \neq \epsilon_m} Re \int \dfrac{d^3k}{(2\pi)^3}\frac{\partial f(\epsilon_n)}{\partial \epsilon_n}\frac{1}{\epsilon_n -\epsilon_m}(v_n^c g_n^{ab}-v_n^ag_n^{cb})
    \end{equation}
The quantity \( g_n^{ab} \) measures the distance between neighboring Bloch states and can be expressed as \( g_n^{ab} = \sum_{ij} \text{Re}[A_{nimj}^a A_{mj,ni}^b] \) for \( n \neq m \). The term \( v_n^c g_n^{ab} \) is known as the quantum metric dipole, where \( A_{nm} = \langle u_n | i \nabla_k u_m \rangle \) represents the Berry connection. In this context,  \( \chi \) is antisymmetric with respect to its first two indices. Due to the time-reversal symmetry of the system, the contribution from the QM is zero because \( \chi_{QM} \) is odd under time reversal.\\
 The contribution due to drude weight is given by,    
\begin{equation}
    \chi_{Drude}^{abc} = -\frac{e^3 \tau^2}{\hbar^3}\sum_n\int \frac{d^k}{(2\pi)^3}(\partial_{k^{a}} \partial_{k^{b}} \partial_{k^{c}}\epsilon_n)f(\epsilon_n)
\end{equation}
In our system, which exhibits time reversal symmetry, the integral becomes an odd function of \( k \), resulting in a total Drude contribution of zero. Therefore, we conclude that the non-linear Hall conductivity arises solely from the BCD.

\section{{CONCLUSION}}
 
The initial excitement surrounding room-temperature ambient-pressure superconductivity in LK-99 has redirected scientific interest toward its unique electronic and topological properties after the inability to reproduce the claimed superconductivity. In this work, we used first-principles density functional theory and developed a refined tight-binding Hamiltonian to investigate the material's band topology and transport characteristics, with a particular focus on its nonlinear Hall response.\\
 Our analysis revealed that, without spin-orbit coupling, two distinct flat bands exist near the Fermi level. The inclusion of spin-orbit coupling leads to the splitting of these bands into four, resulting in multiple Weyl points that act as sources and sinks of Berry curvature. This feature, confirmed by both the augmented and Wannier based Hamiltonians, highlights the topological richness of LK-99. Moreover, the breaking of inversion symmetry in the material gives rise to a non-zero Berry curvature dipole, driving a nonlinear Hall current as a second-order response to the applied electric field. This nonlinear Hall effect, characterized by its quadratic dependence on the electric field and tunability with the field direction, underscores LK-99's potential as a platform for exploring Berry curvature driven transport phenomena. The origin of the nonlinear Hall conductivity is solely from the BCD, as the contributions from the QM and the Drude weight are zero due to time reversal symmetry.\\  
By uncovering these features, our study positions LK-99 as a promising material for advancing the understanding of topological transport and nonlinear electronic devices, independent of its initially proposed superconducting properties.

\acknowledgments
This work was supported by National Science Foundation grant no. DMR 2047193.
   \pagebreak

\appendix

\section{Parameters}
The values of different parameters from the Hamiltonian around the Weyl points are listed in Table \ref{tab1}.\\ 
	$\lambda$=0.0020, $\lambda^{\textquotesingle}$=0.0065\\
\begin{table}[!h ]
  \centering
  \caption{Numerical parameters for the low energy Hamiltonian at different Weyl crossing points}
  \begin{ruledtabular}
\footnotesize
  \begin{tabular}{l l l l l }
  
WP &    a    &  b    &  c & Chern(C) \\
 \hline
   w1 & -0.0065 & -0.044 &  -0.015&   2\\
   w2 & -0.025&   0.256&   0.0045&  -2\\
  
  \end{tabular}
  \end{ruledtabular}
  \label{tab1}
 \end{table}
 \normalsize
 
 \section{Modified tight binding Hamiltonian}
 The parameters here are based on the model by Zhou and Franz \cite{zhou2023cu},
 \begin{equation}
 	H = H_0 + \dfrac{\lambda}{2}\tau_z \otimes \sigma_z + \lambda' \cos(k_zc)\tau_z \otimes \sigma_z 
 \end{equation}

\begin{equation}
H_0 = \begin{bmatrix}
    h_{+}& h_{+-}  \\
    h_{-+} & h_{-} 
        \end{bmatrix}  
\end{equation} 
 
 \begin{equation}
    H = \begin{bmatrix}
    h_{+}+\lambda+\lambda'\cos(k_zc)& h_{+-} & 0 & 0 \\
    h_{-+} & h_{-}-\lambda-\lambda'\cos(k_zc) & 0 & 0 \\
    0 & 0 & h_{+}-\lambda-\lambda'\cos(k_zc) & h_{+-} \\
    0 & 0 & h_{-+} & h_{-}+\lambda+\lambda'\cos(k_zc) 
    \end{bmatrix} 
\end{equation}
 
\begin{align}
     h_{\pm} = E_0 +2t_0 C(k_{||})\pm 2u_0S(k_{||})+2g_0C_z(k_z)\pm 2u_1 S(k_{||}) C_z(k_z) \pm 2 g_{0z} S_z(k_z), \\
     h_{+-} = 2t_1 C_{-}(k_{||}) + 2g_1C_{-}(k_{||})C_z(k_z), \nonumber \\
     h_{-+} = 2t_1 C_{+}(k_{||}) + 2g_1C_{+}(k_{||})C_z(k_z), \nonumber        
\end{align}  

 The basis functions of S3 are,
 
 \begin{align}
    C(k_{||})= \sum_{j=1,2,3} \cos(k_{||}. R_{2j-1})   ,     \\
    S(k_{||})= \sum_{j=1,2,3} \sin(k_{||}. R_{2j-1})   , \nonumber\\
    C_{\pm}(k_{||})= \sum_{j=1,2,3} \omega_{\pm}^{j-1}\cos(k_{||}. R_{2j-1}), \nonumber\\
    C_z(k_z)= cos(k_zc), S_z(k_z)=sin(k_zc)\nonumber
\end{align}  
The values of numerical parameters,\\
$E_0$ = 0.01 ; $t_0$ =- 0.006; $t_1$ = -0.002 ; $u_0$= -0.012; $g_0$ = -0.0125; $u_1$ = 0.001 ; $g_1$ = 0.01; $g_{0z}$ = 0.004

 \section{Effective 2 $\times$ 2 Hamiltonian around different degenerate points}
 \textbf{Around w1}:
\begin{equation}
    H_{w1} = \begin{bmatrix}
    h_{+}-\lambda-\lambda'\cos(k_zc)& h_{+-}  \\
    h_{-+} & h_{-}+\lambda+\lambda'\cos(k_zc) \\
    
    \end{bmatrix} 
\end{equation}

 \textbf{Around w2}:
\begin{equation}
    H_{w2} = \begin{bmatrix}
    h_{+}+\lambda+\lambda'\cos(k_zc)& h_{+-}  \\
    h_{-+} & h_{-}-\lambda-\lambda'\cos(k_zc) \\
    
    \end{bmatrix} 
\end{equation}

\textbf{Around s1}:
\begin{equation}
    H_{s1} = \begin{bmatrix}
    h_{-}-\lambda-\lambda'\cos(k_zc)& 0  \\
    0 & h_{-}+\lambda+\lambda'\cos(k_zc) \\
    
    \end{bmatrix} 
\end{equation}

\textbf{Around s2}:
\begin{equation}
    H_{s2} = \begin{bmatrix}
    h_{+}+\lambda+\lambda'\cos(k_zc)& 0  \\
    0 & h_{+}-\lambda-\lambda'\cos(k_zc) \\
    
    \end{bmatrix} 
\end{equation}
\clearpage

\section{Dispersion of winding vector around Weyl points}
\begin{figure}[!h]
    \centering
    \includegraphics[width=0.9\linewidth]{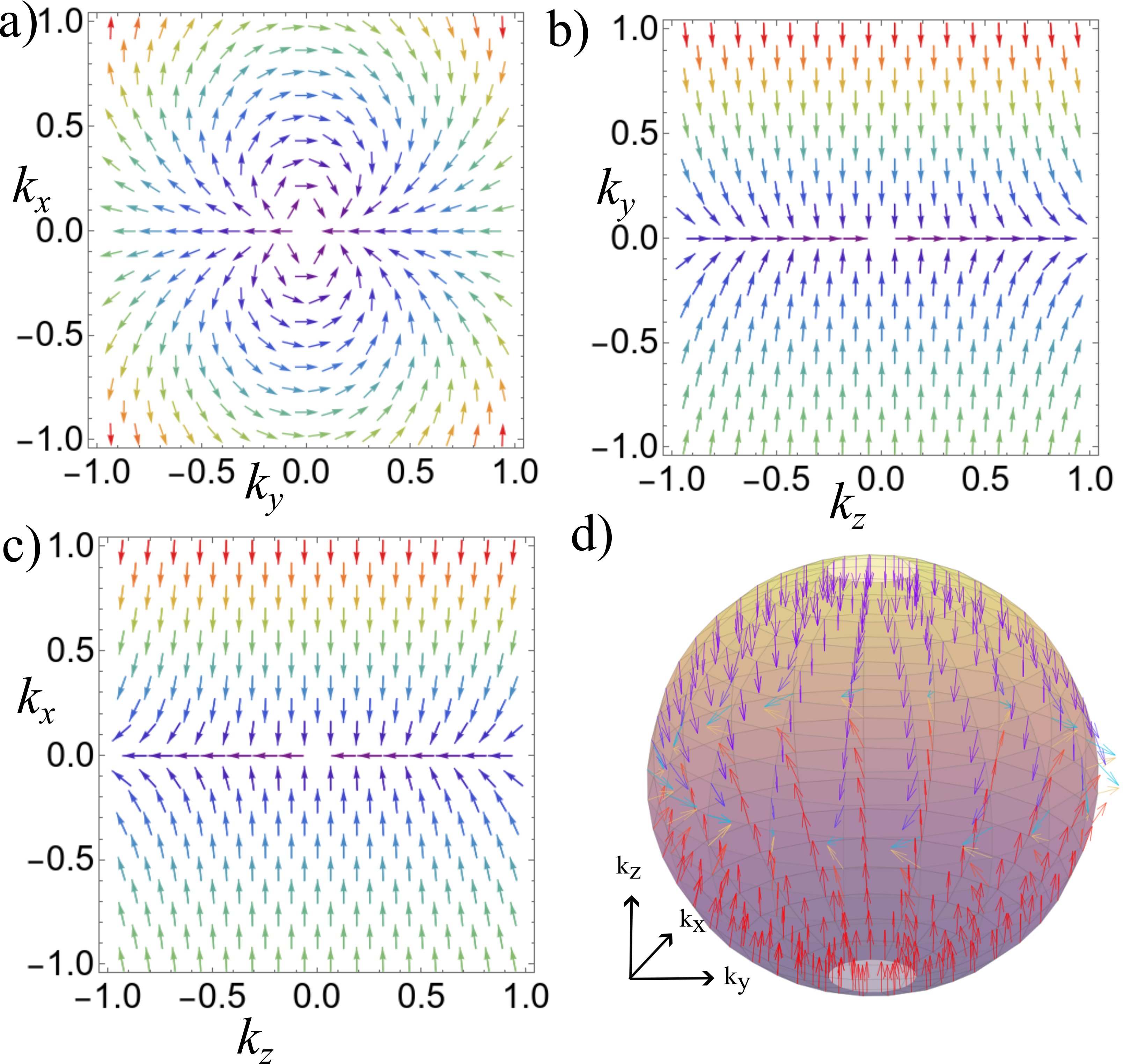}
    \caption{Visualization of the  winding textures of the vector field $\mathbf{f}(\mathbf{k})$ in momentum space. 
(a) Winding texture in the $k_x$–$k_y$ plane at fixed $k_z = 0$. 
(b) Texture in the $k_y$–$k_z$ plane at fixed $k_x = 0$. 
(c) Texture in the $k_x$–$k_z$ plane at fixed $k_y = 0$. 
(d) Texture on the unit sphere, illustrating the full 3D structure of $\mathbf{f}(\mathbf{k})$. The arrows show the direction of $\hat{\mathbf{f}}(\mathbf{k})$. The winding structure reveals the topological nature of the Hamiltonian.}
    \label{fig:winding}
\end{figure}
\clearpage

\section{Berry Curvature}
The Hamiltonian at the Weyl points (w1 and w2) is given by the equation,
\begin{equation}
    H = a(k_x^2-k_y^2)\tau_x + 2ak_xk_y\tau_y+(bk_z+ck_z^2)\tau_z
\end{equation}

The eigen values obtained analytically,
\begin{equation}
    \lambda = \pm\sqrt{a^2(k_x^2-k_y^2)+(2ak_xk_y)^2+(bk_z+ck_z^2)}
\end{equation}
The band velocities can be simply obtained by ;
\begin{equation}
    v_i=\frac{\partial \lambda }{\partial k_i}
\end{equation}
The different components of Berry curvature can be obtained by simply using the formula:
\begin{equation}
    \Omega=\frac{1}{2}\frac{\vec{f}}{|\vec{f}|}.\left(\frac{\partial \vec{f}}{\partial k_j}\times \frac{\partial \vec{f}}{\partial k_k}\right)\hat\epsilon_i
    \end{equation}
    
Different components of $\vec{f}$ are; $f_x = a(k_x^2-k_y^2)$; $f_y=2ak_xk_y$; $f_z=bk_z+ck_z^2$\\
Using eqn (4), the different components of Berry curvature are:\\
$$\Omega_z=\frac{4a^2(bk_z+ck_z^2)(k_x^2+k_y^2)}{[a^2(k_x^2+k_y^2)^2+(bk_z+ck_z^2)^2]^{3/2}}$$
$$\Omega_x=\frac{4a^2(bk_z+ck_z^2)(k_y^2-k_x^2)}{[a^2(k_y^2+k_x^2)^2+(bk_z+ck_z^2)^2]^{3/2}}$$
$$\Omega_y=\frac{8a^2(bk_z+ck_z^2)(k_xk_y)}{[a^2(k_x^2+k_y^2)^2+(bk_z+ck_z^2)^2]^{3/2}}$$

With simply using Berry curvature and band velocity one can calculate the Berry curvature dipole $D_{ij}$  $\propto$ $\sum_n v_i^n \Omega_j^n$ 

\clearpage

\section{Temperature Dependence of BCD}
\begin{figure}[!h]
    \centering
    \includegraphics[width=0.7\linewidth]{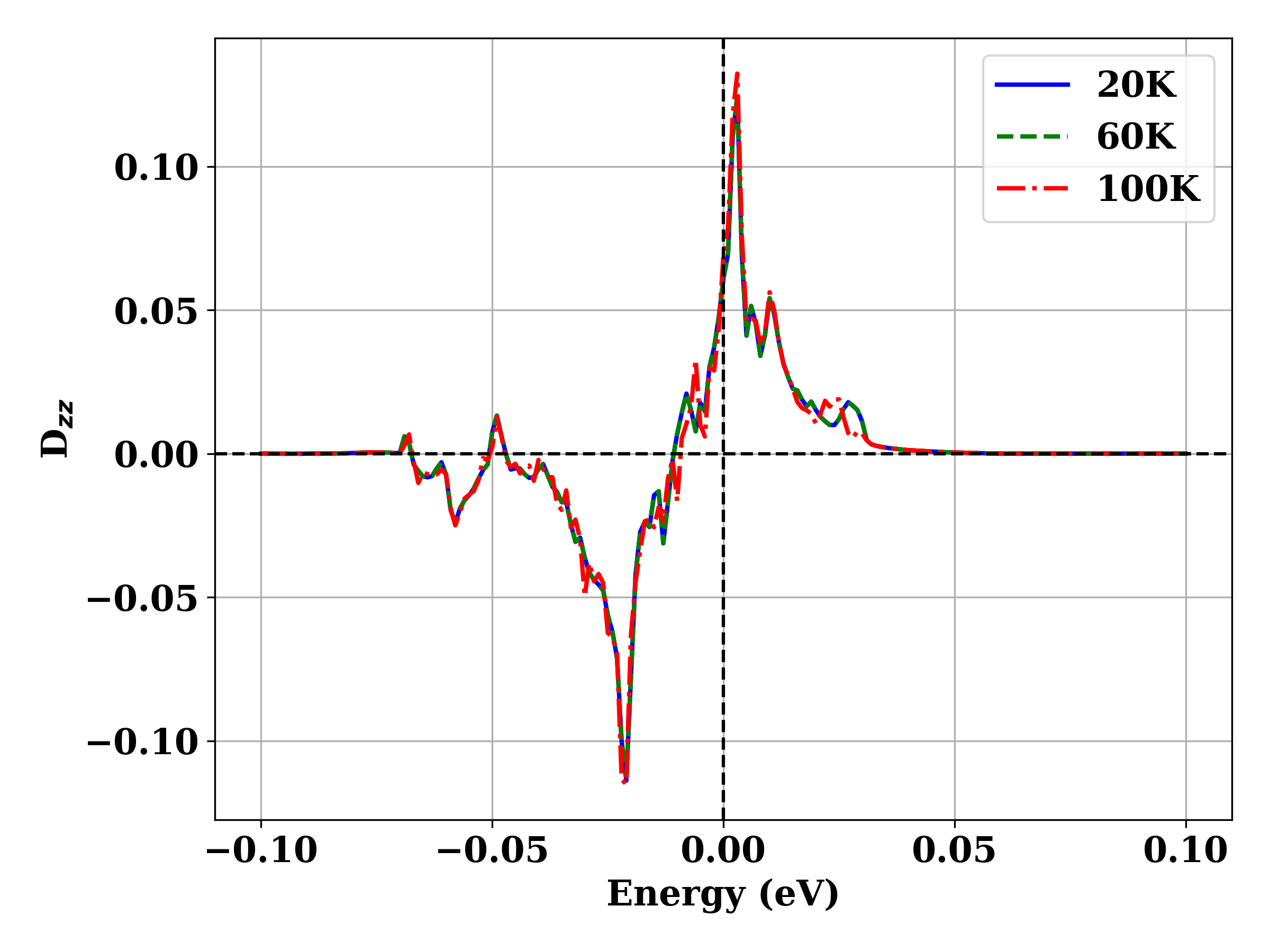}
    \caption{Temperature dependence on one of the component($D_{zz}$) of Berry Curvature Dipole}
    \label{fig:qm}
\end{figure}
\clearpage

\section{Wannier Function Model Bandstructure}
\begin{figure}[!h]
   \centering
   \includegraphics[scale=0.4]{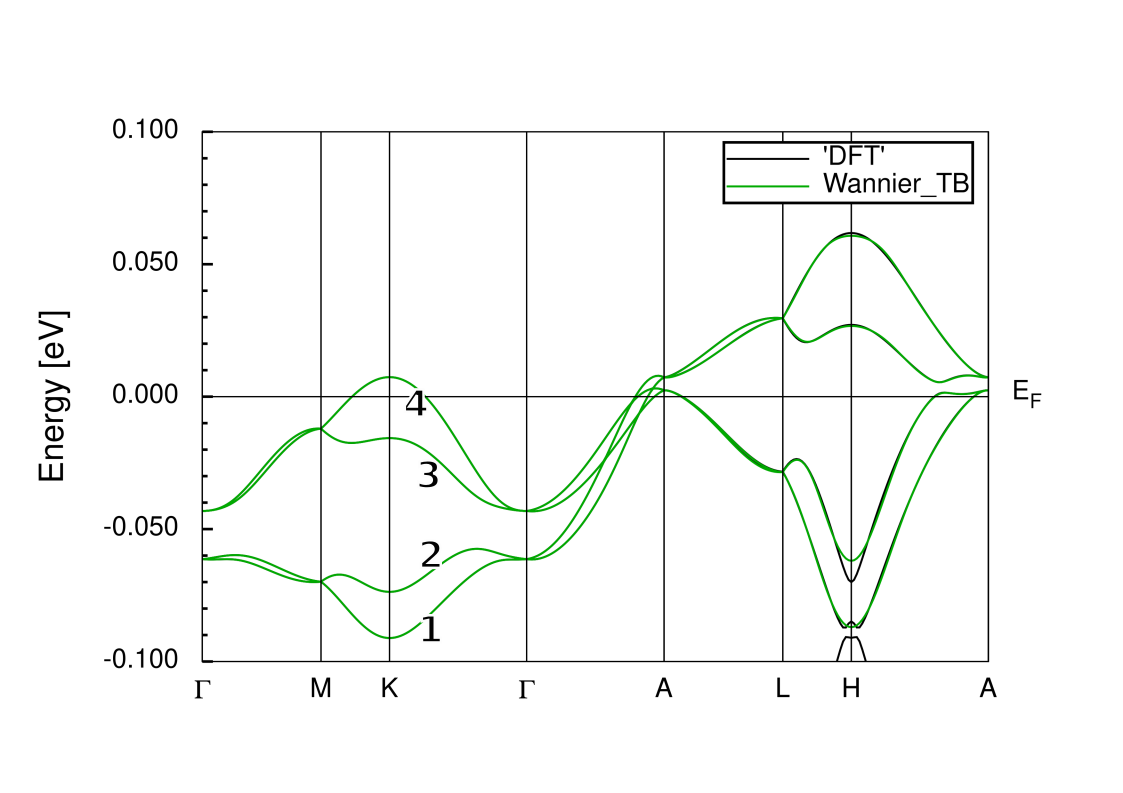}
   \caption{Comparision of band structures from DFT calculations(black) and Wannier function model (green) .}
    \label{fig6}
  \end{figure}

\section{Distribution of Berry Curvature}
 \begin{figure}[!h]
   \centering
   \includegraphics[width=0.9\columnwidth]{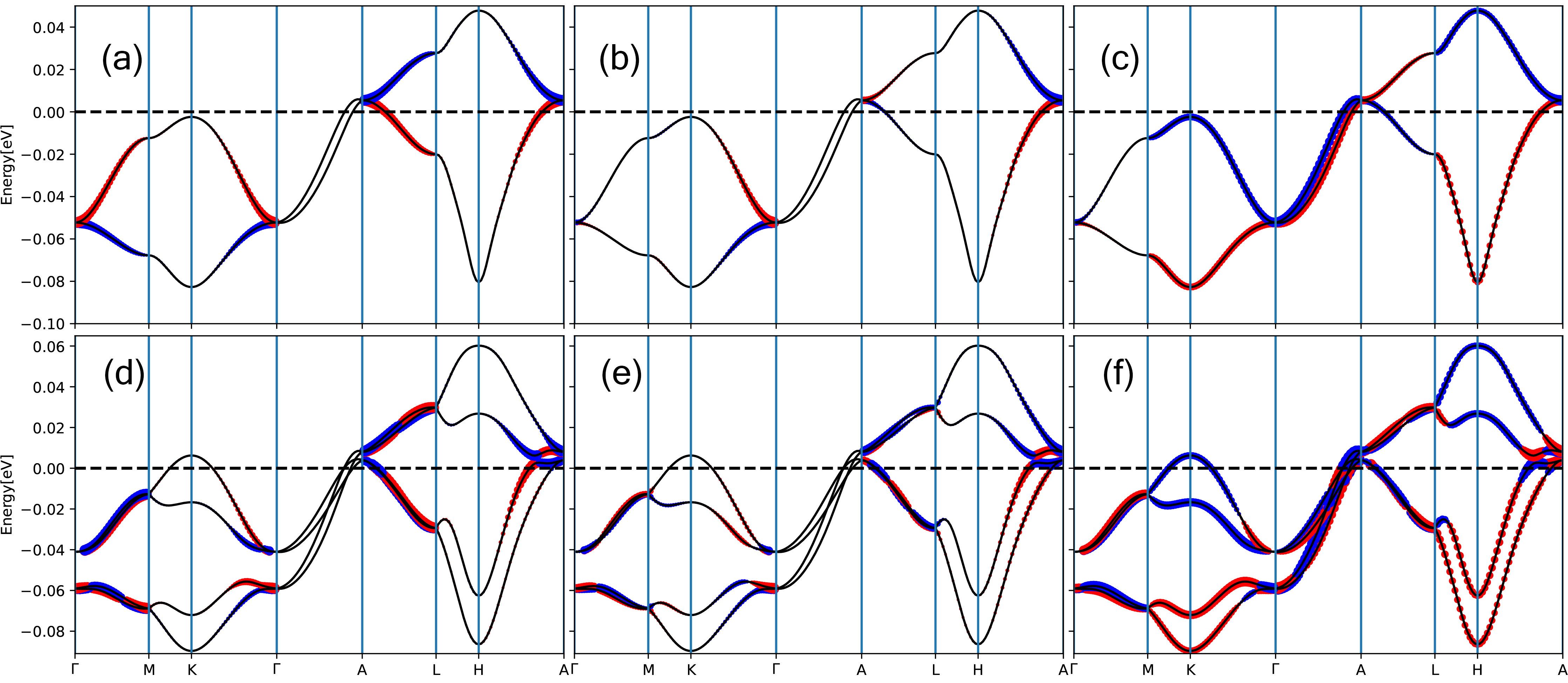}
   \caption{Projection of the components of Berry curvature on the individual bands. The first three are  (a) $\Omega_x$ (b) $\Omega_y$ and (c) $\Omega_z$ without SOC. The next three are the case with SOC (d) $\Omega_x$ (e) $\Omega_y$ and (f) $\Omega_z$. Red and blue colors represent the positive and negative values of the Berry curvature. The thickness indicates the magnitude of Berry curvature.}
    \label{fig7}
  \end{figure}
\section{Weyl Points}
\subsection{Absence of SOC}
\begin{table}[!h]
\centering
 \caption{Properties of low-energy Weyl points for two bands without SOC. "WPs" represents the set of Weyl points, $E$ indicates their energy relative to the Fermi level, $k_i$ $(i = x, y, z)$ are their coordinates, "M" denotes the symmetry-determined multiplicity of the Weyl points, and $\chi$ refers to their associated chirality.}
 \footnotesize
\begin{ruledtabular}
\begin{tabular}{l l l l l l l}

WP & $k_x[2 \pi / a]$ & $k_y [2 \pi / b]$ & $k_z [2 \pi / c]$ & $E (eV)$ & $\chi$ & M \\
\hline
W$_{+}$ & 0.000 & 0.000 & 0.000 & -0.052 & +1 & 1 \\
W$_{-}$ & 0.000 & 0.000 & -0.688 & 0.005 & -1 & 1 \\

\end{tabular}
\end{ruledtabular}
\label{tab:wave-vectors}
\end{table}
\normalsize

\begin{figure}[!h]
   \centering
   \includegraphics[scale=0.1]{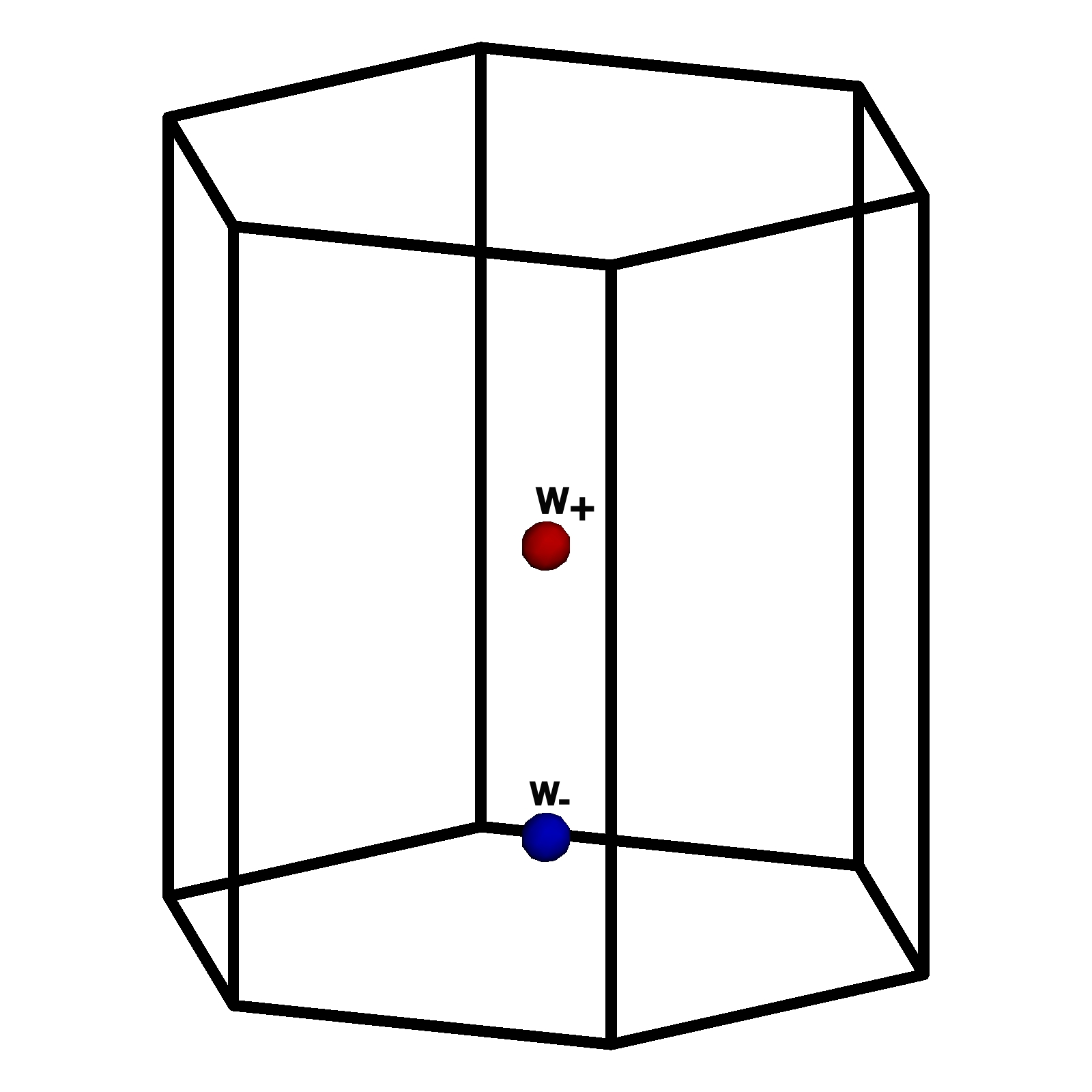}
   \caption{Two Weyl points distributed in the 3D Brillouin zone for the case of absence of SOC. The red and blue colors indicate the positive and negative chirality respectively.}
    \label{fig8}
  \end{figure}

\subsection{Inclusion of SOC}
\begin{table}[!h]
\centering
 \caption{Properties of low-energy Weyl points in LK-99, considering the bands $n=1$ and $n+1=2$. "WPs" represents the set of Weyl points, $E$ denotes their energy relative to the Fermi level, $k_i$ $(i = x, y, z)$ are their coordinates, "M" refers to the symmetry-determined multiplicity of these Weyl points, and $\chi$ indicates their associated chirality.}
\begin{ruledtabular}
\footnotesize
\begin{tabular}{l l l l l l l}

WP & $k_x[2 \pi / a]$ & $k_y [2 \pi / b]$ & $k_z [2 \pi / c]$ & $E (eV)$ & $\chi$ & M \\
\hline
W$_1$ & -0.867 & -0.500 & 0.000 & -0.069 & -1 & 3 \\
W$_2$ & -0.506 & -0.012 & -0.097 & -0.065 & +1 & 6 \\
W$_3$ & -0.866 & -0.500 & -0.668 & -0.029 & -1 & 3 \\
W$_4$ & -0.163 & -0.003 & 0.668 & -0.003 & +1 & 6 \\
W$_5$ & -0.120 & 0.062 & -0.173& -0.049 & -1 & 6 \\
W$_6$ & 0.000 & 0.000 & -0.668 & 0.004 & -1 & 1 \\
W$_7$ & 0.000 & 0.000 & 0.000 & -0.059& +1 & 1 \\
W$_8$ & 0.000 & 0.000 & 0.411 & -0.004 & +1 & 2 \\

\end{tabular}
\end{ruledtabular}
\label{tab:wave-vectors}
\end{table}
\normalsize

\begin{figure}[!h]
   \includegraphics[width=\columnwidth]{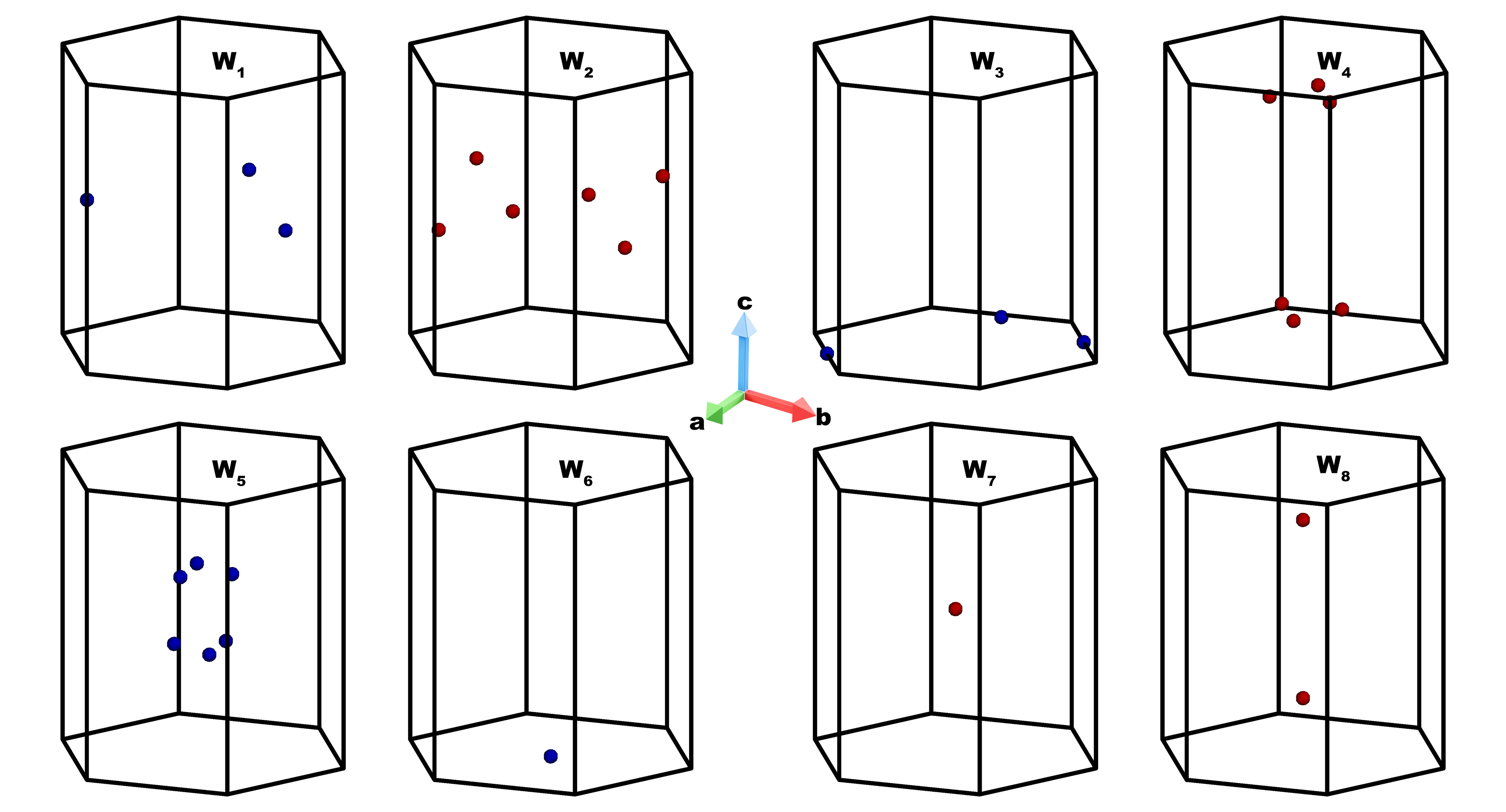}
   \caption{Eight sets of Weyl points distribution in the 3D Brillouin zone. The red and blue colors indicate the opposite Chern number containing Weyl points.}
    \label{fig9}
  \end{figure}

\begin{table}[h]
\centering
  \caption{Properties of low-energy Weyl points in LK-99, considering the bands $n=2$ and $n+1=3$. "WPs" represents the set of Weyl points, $E$ denotes their energy relative to the Fermi level, $k_i$ $(i = x, y, z)$ are their coordinates, "M" refers to the symmetry-determined multiplicity of these Weyl points, and $\chi$ indicates their associated chirality.}
  \footnotesize
\begin{ruledtabular}
\begin{tabular}{l l l l l l l }

WP & $k_x[2 \pi / a]$ & $k_y [2 \pi / b]$ & $k_z [2 \pi / c]$ & $E (eV)$ & $\chi$ & M \\
\hline

W$_9$ & -0.210 & -0.097& -0.285& -0.021 & +1 & 6 \\
W$_{10}$ & -0.185 & 0.088 & -0.461 & 0.003 & -1 & 6 \\
W$_{11}$ & -0.000 & 0.000 & 0.301 & -0.019 & +1 & 2 \\
W$_{12}$ & 0.000 & -0.000 & 0.459 & 0.004 & -1 & 2 \\

\end{tabular}
\end{ruledtabular}
\end{table}
\normalsize

 \begin{figure}[!h]
   \includegraphics[width=0.9\columnwidth]{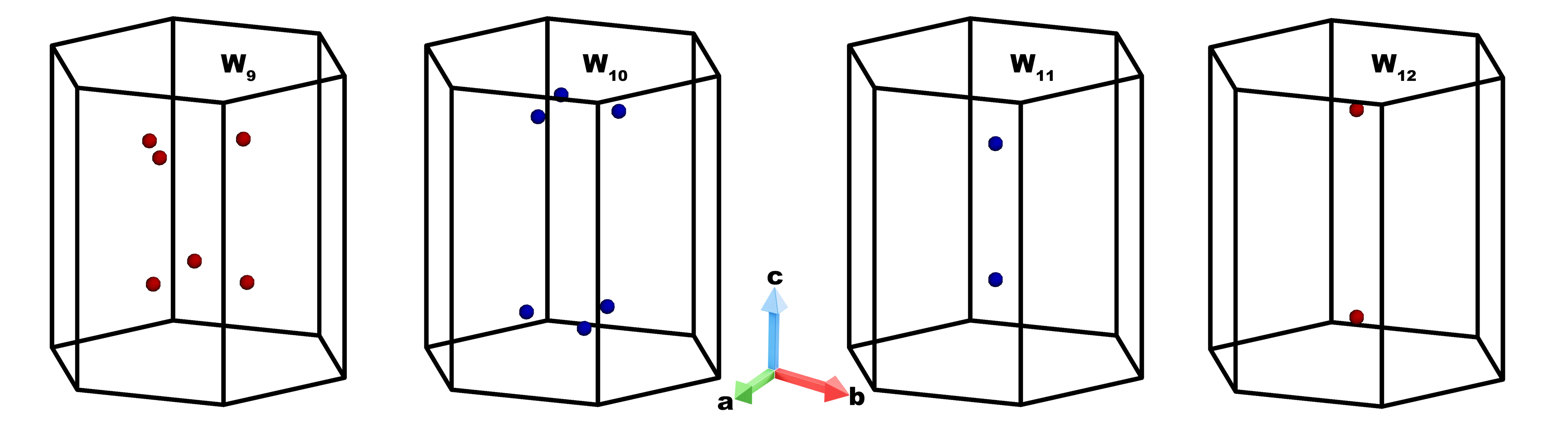}
   \caption{Eight sets of Weyl points distribution in the 3D Brillouin zone. The red and blue colors indicate the opposite Chern number containing Weyl points.}
    \label{fig10}
  \end{figure}

\begin{table}[!h]
\centering

 \caption{Properties of low-energy Weyl points in LK-99, considering the bands $n=3$ and $n+1=4$. "WPs" represents the set of Weyl points, $E$ denotes their energy relative to the Fermi level, $k_i$ $(i = x, y, z)$ are their coordinates, "M" refers to the symmetry-determined multiplicity of these Weyl points, and $\chi$ indicates their associated chirality.}
 \footnotesize
\begin{ruledtabular}
\begin{tabular}{ l l l l l l l }

WP & $k_x[2 \pi / a]$ & $k_y [2 \pi / b]$ & $k_z [2 \pi / c]$ & $E (eV)$ & $\chi$ & M \\
\hline

W$_{13}$ & -0.577 & 0.000 & -0.688 & 0.030 & -1 & 3 \\
W$_{14}$ & -0.866 & -0.500 & 0.000 & -0.013 & +1 & 3 \\
W$_{15}$ & -0.158 & 0.0152 & 0.354 & -0.000 & +1 & 6 \\
W$_{16}$ & -0.000 & -0.000 & 0.000 & -0.041 & -1 & 1 \\
W$_{17}$ & -0.000 & 0.000 & 0.392 & -0.000 & -1 & 2 \\
W$_{18}$ & 0.000 & 0.000 & -0.688 & 0.008 & -1 & 1 \\

\end{tabular}
\end{ruledtabular}
\end{table}

\normalsize

  \begin{figure}[!h]
   \centering
   \includegraphics[width=0.7\columnwidth]{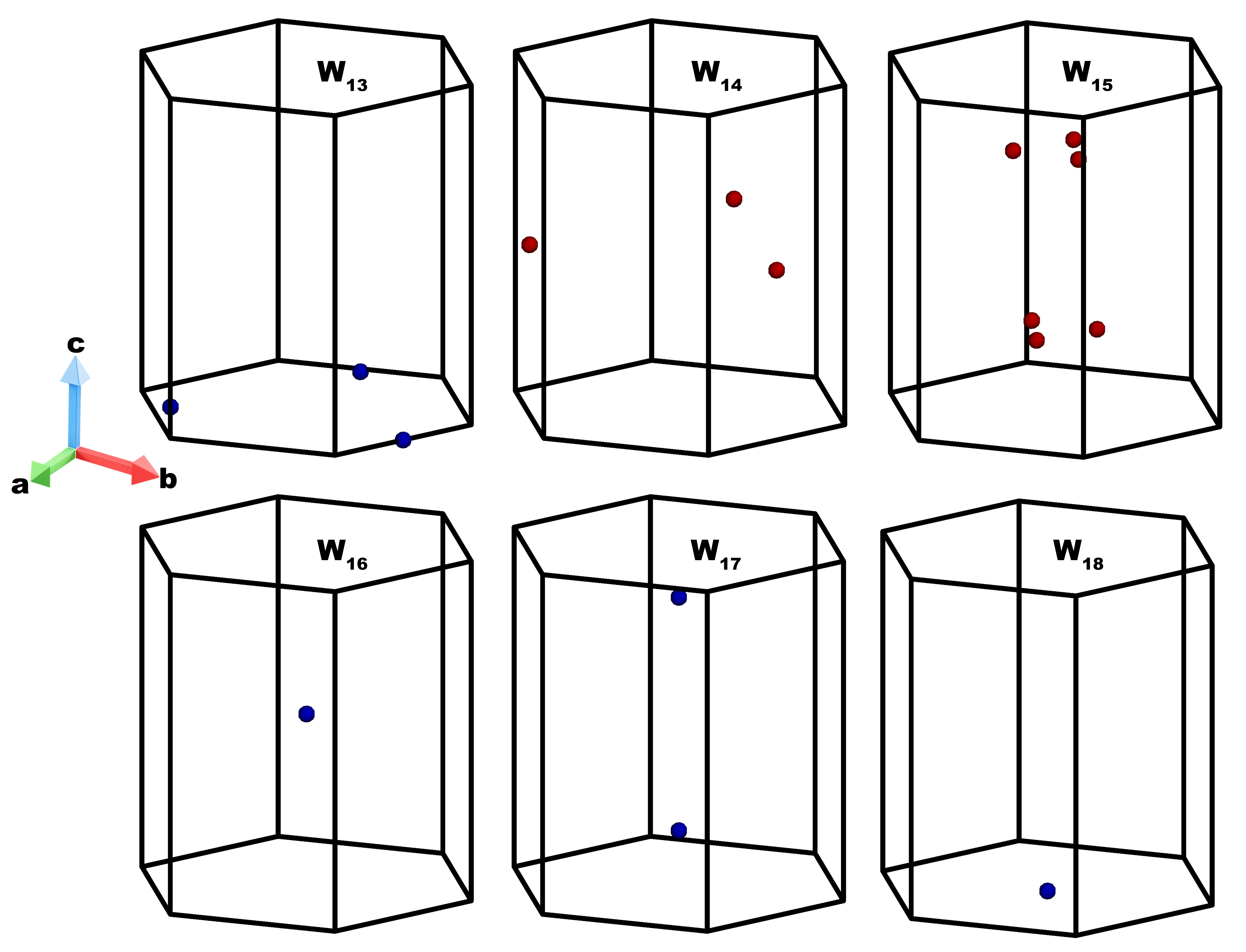}
   \caption{Eight sets of Weyl points distribution in the 3D Brillouin zone. The red and blue colors indicate the opposite Chern number containing Weyl points.}
    \label{fig11}
  \end{figure}

\clearpage
\bibliography{ref}
\end{document}